%% file: scaling.tex
\documentstyle[psfig]{europhys}
\input euromacr

\begin{document}
\euro{}{}{}{}
\Date{}
\shorttitle{}
\title{
Nonlinear response and scaling law  in
the vortex state of $d$-wave superconductors}
\author{H. Won\inst{1,2}
\And K. Maki\inst{2}}
\institute{
\inst{1}Department of Physics, Hallym University,
Chunchon 200-702, South Korea \\
\inst{2}Department of Physics and Astronomy,
 University of Southern California, Los Angeles, CA 90089-0484,
USA}               
\rec{}{}
\maketitle
\begin{abstract} 
We study the field dependence of the quasi-particle density
of states,
the thermodynamics and the transport properties in the
vortex state of $d$-wave superconductors when a magnetic field
is applied 
perpendicular to the conducting plane,  specially for
the low field  and the low temperature compared to the upper critical field
and transition temperature, respectively,
$H/H_{c2} \ll 1$ and $T/T_c \ll 1$.
Both the superfluid density and the spin susceptibility exhibit
the characteristic $\sqrt{H}$-field dependence, while the nuclear
spin lattice relaxation rate T$_1^{-1}$ and the thermal
conductivity are linear in field $H$. With increasing temperature,
these quantities exhibit the scaling behavior in $T/\sqrt{H}$.
The present theory applies to 2D $f$-wave superconductor as well;
a possible candidate of the superconductivity in Sr$_2$RuO$_4$. 
\end{abstract}

\section{Introduction}
Single crystals of high-$T_c$ cuprate superconductors like
YBCO, Bi2212, etc. appear to provide the most useful testing ground
for properties of unconventional superconductors\cite{maki_ann}. In
particular $d$-wave superconductivity has been established both in the
hole-doped and the electron-doped high-$T_c$ 
cuprates \cite{tsuei,kokales,prozorov}. 
Due to the nodal
structure in the $d$-wave  order parameter,  the specific heat\cite{volovik}, 
the spin
susceptibility and the superfluid density\cite{won_euro} have been predicted
to behave like $\sqrt{H}$ in the vortex state where $H$ is the magnetic
field. Indeed the $\sqrt{H}$ dependence of the specific heat in the vortex
state of YBCO has been established \cite{moler,wright,revas}. 
On the other hand, the 
$\sqrt{H}$
dependence of the superfluid density appears to 
have not been seen
in spite of an elaborate muon spin rotation experiment \cite{sonier}.
\par
In the meantime, the calculational technique greatly 
improved \cite{barash,kubert,vekhter,won_nato}.
Therefore we can study the effect of a magnetic field within the
semi-classical approximation almost analytically within the
weak-coupling theory of $d$-wave superconductivity. 
Most of earlier works considered only the spherical Fermi surface
\cite{barash,kubert,vekhter}.
Here
we assume that the Fermi surface of these $d$-wave
superconductors is a weakly modulated cylinder\cite{won_nato}. Then the
present model should be also applicable to the recently discovered
$d$-wave superconductors in $\kappa$-(ET)$_2$ salts 
\cite{carrington,pinteric,ichimura}.
The present model applies as well to 
the superconductivity in  
Sr$_2$RuO$_4$,
if the superconductivity is
one of 2D $f$-wave states
(i.e.
$\Delta({\bf k}) \sim e^{\pm i \phi}\cos(2\phi)$,
$e^{\pm i \phi}\sin(2\phi)$ or
$e^{\pm i \phi}\cos(ck_3)$,
where $\phi$ is the angle between the
quasi-particle  momentum 
and $a$-axis in the plane and $k_3$ the 
quasi-particle momentum
in the $c$-direction and $c$ is the distance 
between layers
of conducting plane in the $c$-direction.)
which are considered by Hasegawa {\it et al.} \cite{hasegawa},
The specific heat \cite{nishizaki}.
NMR \cite{ishida} and 
the magnetic penetration depth measurement \cite{bonalde}
show clearly  the presence of the nodal structure \cite{won_press}.
The thermodynamics and the planar transport of these $f$-wave states
are exactly same as the ones 
in $d$-wave superconductors when the field is applied 
perpendicular to the conducting plane.
\par
The object of the present paper is to study the quasi-particle
density of states in the vortex state of $d$-wave superconductors
when a magnetic field
is applied
perpendicular to the conducting plane,  specially for
the low field and the low temperature compared to the upper critical field
and transition temperature, respectively,
$H/H_{c2} \ll 1$ and $T/T_c \ll 1$.
Then
making use of the density of states, we calculate the specific
heat, the spin susceptibility, the superfluid density and 
the nuclear spin lattice relaxation rate T$_1^{-1}$ in
NMR. Also for clarity, we limit ourselves to the superclean limit
where the effect of impurity scattering is negligible.
For $E/\Delta
\ll 1$  the
quasi-particle density of states is a simple function of 
$E/\epsilon \sim E/\sqrt{H}$ 
where $E$ is the energy of a quasi-particle,  $\Delta$ is the $d$-wave 
superconducting order parameter, 
$\epsilon =  v\sqrt{eH}/2$, $e$ electron charge and $v$ the Fermi velocity
in the conducting plane. Then in the limit of  $T \rightarrow 0$, 
the specific heat, the spin
susceptibility and the change in the superfluid density behave
like $\sqrt{H}$, while T$_1^{-1}$ behaves like $H$. Indeed, the $\sqrt{H}$
dependence of the
spin susceptibility and the $H$ linear dependence of T$_1^{-1}$, are
recently observed by NMR in the vortex state of underdoped
YBCO \cite{zheng}.
In addition, the thermal conductivity 
both in the superclean limit
and the clean limit
 will be briefly discussed.
\par
For $T \neq 0$ K, all these quantities exhibit scaling behavior as first
discussed by Simon and Lee \cite{simon}.  Actually the present model gives
the explicit scaling functions, which should be readily
accessible experimentally.

\section{Quasi-particle density of states}
Following the semi-classical approximation by Volovik\cite{volovik}
 the quasi-particle density of states in the
the vortex state of $d$-wave superconductor is given by
\begin{equation}
N(E,H)/N_0 \simeq \frac1\Delta 
\langle |E| \vee 
|{\bf v} \cdot {\bf q}| \rangle
\end{equation}
where $ |E| \vee
|{\bf v} \cdot {\bf q}|$ means the bigger one
among  $|E|$ and $|{\bf v} \cdot {\bf q}|$.
Also we assumed $|E|$,  $|{\bf v} \cdot {\bf q}|
\ll \Delta$. Here $|{\bf v} \cdot {\bf q}|$ is the Doppler shift
\cite{maki_prog}
associated the pair momentum $2{\bf q}$ and 
the Fermi velocity $\bf v$.
In a magnetic field 
${\bf H} \parallel {\bf c}$, the Doppler shift
is given as ${\bf v} \cdot {\bf q}$
=$\displaystyle \frac{v}{2r}\cos\phi$ 
where $r$ is the distance from the center
of the vortex  and $\phi$ is the angle 
between ${\bf v}$ and 
$\bf q $.
Although  ${\bf v}$ is parallel
to one of nodal lines  at low temperature,
$\phi$ runs from  0 to $2\pi$.
Finally $\langle \ldots \rangle$
of the Eq.(1) means 
 the spatial average over $r$
and  $\phi$.
This
average is carried out over a unit cell of a square
vortex lattice characteristic to $d$-wave superconductors 
\cite{won_euro} $\acute{a}$ la 
Wigner Seitz.
\begin{equation}
\langle \ldots \rangle
=\frac1{2\pi} \int_0^{2\pi}d\phi 
\frac2{d^2}\int_0^d r dr \ldots
\end{equation}
where 2$d$ is the distance between vortices, $d=1/ \sqrt{eH}$ \cite{won_nato}.
Then Eq.(1) reduces to 
\begin{equation}
N(E,H)/N_0 =
\frac4\pi \frac{\epsilon}{\Delta}
g(\frac{E}{\epsilon}) 
\simeq\frac{2 v}{\pi}\sqrt{\frac{\pi}{\Phi_0}} \frac{\sqrt{H}}{\Delta}
g(\frac{2}{v}\sqrt{\frac{\Phi_0}{\pi}} \frac{E}{\sqrt{H}})
\end{equation}
with the scaling function $g(s)$,

\begin{equation}
g(s) =\left\{ 
\begin{array}{ll}
  \frac\pi4 s(1+ \frac1{2 s^2}) 
&  \mbox{for  
$s=\frac{E}\epsilon \ge 1$} \\
 \frac34 \sqrt{1-s^2}
+ \frac{1}{4s}(1+2 s^2)\sin^{-1}s
&
\mbox{for $ s=E/\epsilon \le 1$}
\end{array}
\right.
\end{equation}
and $\displaystyle \epsilon= \frac{v}{2}\sqrt{eH} 
=\frac{v}{2}\sqrt{\frac{\pi}{\Phi_0}} 
\sqrt{H}$
and $\Phi_0$ is a quantum of flux 
($\simeq 2.07 \times 10^{-11}$ T-cm$^2$).
For $E=$ 0, we obtain $N(0,H)/N_0 = 
\frac4\pi \frac{\epsilon}{\Delta}$,
since $g(s)= 1 + \frac16 s^2 $ for $s \ll 1$. 
Also
$g(s) \sim \frac{\pi}{4}s$    for $s \gg1$ which means 
$N(E,H)/N_0 \simeq E /\Delta $ for the high energy excitations 
$E \gg \epsilon$.
The scaling function $g(s)$ versus 
$s=\frac{E}\epsilon(= \frac{2}{v} \sqrt{\frac{\Phi_0}{\pi}} 
\frac{E}{\sqrt{H}})$
is shown in Fig. 1.
A function similar to Eq.(4) has been obtained for a spherical
Fermi surface in \cite{kubert}.

\begin{figure}
\caption{ The scaling function $g(s)$ in Eq.(4) is shown as
a function of $s=E/\epsilon$.  
}
\end{figure}

\section{Thermodynamics}
Making use of the density of states, i.e. Eq.(3),
we can determine the thermodynamic quantities
at the low temperatures, $T \ll T_c$. 
First, the specific heat $C_s(T,H)$ is given by
\begin{eqnarray}
C_s(T,H)/\gamma_n T 
&=& \frac{3}{2\pi^2}
\int_0^{\infty}  d E E^2 
\frac{N(E,H)}{N_0}
{\rm sech}^2(\frac{E}{2T})
\nonumber\\
&=& \frac4\pi \frac\epsilon\Delta f(T/\epsilon)
\end{eqnarray}
where 
\begin{equation}
f(T/\epsilon)=
\frac3{2\pi^2}
(\frac{\epsilon}{T})^3
\int_0^{\infty}ds s^2g(s)
{\rm sech}^2(\frac{\epsilon}{2T}s) 
\end{equation}
and $\gamma_n T$ is the specific heat in the normal state.
We have $\displaystyle f(T/\epsilon) \rightarrow 1$ 
for $T/\epsilon \ll 1$ 
and $\displaystyle f(T/\epsilon)=\frac{27\zeta(3)}{4\pi}
\frac{T}\epsilon$
for $T/\epsilon \gg 1$.
With this limiting behavior of $f(T/\epsilon)$,
the specific heat takes  
$C_s/\gamma_n T
=\frac4\pi \frac{\epsilon}{\Delta}
= \frac{2 v}{\pi}\sqrt{\frac{\pi}{\Phi_0}} 
\frac{\sqrt{H}}{\Delta}$
at low temperature
$ T \ll \epsilon$
and $C_s/\gamma_n T = \displaystyle
\frac{27\zeta(3)}{\pi^2} \frac{T}{\Delta}$
for the $H=0$, respectively.
In the limit of $T \rightarrow 0$, the $\sqrt{H}$ dependence of
the specific heat in YBCO  has been discussed in \cite{moler,wright,revas}.
When we parameterized 
$C_s/T= A_c \sqrt{H}$  and  the specific heat in the absence of
the field $C_s(T,0)/T = \alpha T$,
we can deduce 
$\displaystyle  
v = \frac{27 \zeta(3)}{2\pi}
\sqrt{\frac{\pi}{\Phi_0}} \frac{A_c}{\alpha} \simeq 
2.28 \times 10^6$ cm/s  
and $1.76 \times 10^6$ cm/s from \cite{moler} and 
\cite{revas}, 
respectively.
We may introduce an adjustable parameter $a$ in front of $v$
as in 
\cite{kubert}.
Then if we assume 
the Fermi velocity  $v\simeq 10^7$ cm/s,
$a$ is about  0.2
which is the same as deduced by Chiao {\it et al.}\cite{chiao}.

\begin{figure}
\caption{The specific heat data ($\Diamond$) by
Nishizaki {\it et al.}[19] is fitted 
with the  } 
$\sqrt H$  law.
\end{figure}

\par
In the case of Sr$_2$RuO$_4$ the recent specific heat data by
Nishizaki {\it et al.}\cite{nishizaki}
exhibits clearly the  $\sqrt{H}$ dependence
as shown in Fig.2.
The deviation from the  $\sqrt{H}$ law for $H < 0.01 T$ 
is most likely
due to disorders.
\par
Also Wang {\it et al.} \cite{wang} studied the 
scaling behavior of the specific heat of YBCO
\begin{eqnarray}
\frac{C(T,H)-C(T,0)}
{\gamma_n T  \,\,\,\,[\displaystyle
\frac{4}{\pi}
 \frac{\epsilon}{\Delta}] }
&\equiv& F(\frac{T}{\epsilon})
 = f(\frac{T}{\epsilon})
- \frac{27 \zeta(3)}{4\pi}
 \frac{T}{\epsilon} \nonumber\\
&=& \left\{
\begin{array}{ll}
1   -\frac{27\zeta(3)}{4\pi} \frac{T}{\epsilon}
\frac{\pi^2}{30}(\frac{T}{\epsilon})^2 + \cdots
& \mbox{for $\frac{T}{\epsilon} \ll 1$} \\
\frac{3}{2\pi} (\ln2) (\frac{\epsilon}{T})^2 + \cdots 
&
\mbox{for $\frac{T}{\epsilon} \gg 1$}
\end{array}
\right.
\end{eqnarray}
We show in Fig.3 $F(\frac{T}{\epsilon})$
 versus $\frac{T}\epsilon$.
The present model appears to describe the scaling behavior
in YBCO \cite{wang} very well.
Clearly a similar study of the scaling relation in the vortex state of 
Sr$_2$RuO$_4$ will be of great interest.

\begin{figure}
\caption{
The
scaling function $F(T/\epsilon)$
of specific heat in Eq.(7), 
the scaling function $I(T/\epsilon)$
 of the spin susceptibility
(or the superfluid density) in Eq.(10)
and the scaling function $G(T/\epsilon)$
of the nuclear spin lattice relaxation rate in Eq.(13)
 are shown altogether as a function of $T/\epsilon$.
}
\end{figure}

\par
Similarly, both the spin
susceptibility $\chi$ and the superfluid density $\rho_s$
should exhibit the scaling behavior.
\begin{eqnarray}
\chi(T,H)/\chi_n   
&=& 1- \rho_s(H,T)
=  \frac{1}{2T}
\int_0^{\infty} 
\frac{N(E, H)}{N_0}
{\rm sech}^2(\frac{E}{2T})
\nonumber\\
&=& \frac4\pi \frac\epsilon\Delta h(T/\epsilon)
\end{eqnarray}  
and \begin{equation}
h(T/\epsilon)=
\frac12
\frac{\epsilon}{T}
\int_0^{\infty}ds g(s)
{\rm sech}^2(\frac{\epsilon}{2T}s)
\end{equation} 
Here $\chi_n$ is the spin
susceptibility in the normal state.
We have
$\displaystyle h(T/\epsilon) \rightarrow 1$
for $T/\epsilon \ll 1$
and $\displaystyle h(T/\epsilon)=
\frac\pi2\ln 2 \frac{T}\epsilon$
for $T/\epsilon \gg 1$.
With this limiting behavior of $h(T/\epsilon)$
both the susceptibility and superfluid density  
take $\sim  \frac{2v}{\pi}\sqrt{\frac{\pi}{\Phi_0}}\frac{\sqrt{H}}{\Delta}$
in the limit of  $T \rightarrow 0$, 
and $ \sim 2 (\ln2) \frac{T}{\Delta}$
in the limit of $ T \gg \epsilon$
As already mentioned in the introduction the 
$\sqrt{H}$dependence of the susceptibility has been observed \cite{zheng}.
However the reported $\rho_s(T,H)$ in YBCO by muon spin rotation
appears not to exhibit the $\sqrt{H}$ behavior.  
A further study of $\rho_s(T,H)$
is  clearly desirable.
Again it may be more useful to introduce the scaling function by
\begin{eqnarray}
I(T/\epsilon) \equiv 
\frac{\chi(T,H) - \chi(T,0)}
{\chi_n [\frac4\pi \frac{\epsilon}{\Delta}]}
= h(T/\epsilon) - 2 \ln 2 (T/\epsilon)
\nonumber\\
\simeq
\left\{
\begin{array}{ll}
1   -2 \ln 2 \frac{T}{\epsilon} + \cdots 
& \mbox{for $\frac{T}{\epsilon} \ll 1$} \\
\frac{\pi}{16} \ln(1.622 \frac{T}{\epsilon}) \frac{\epsilon}{T}
&
\mbox{for $\frac{T}{\epsilon} \gg 1$} 
\end{array} 
\right.
\end{eqnarray}
The scaling function $I(T/\epsilon)$ versus
$T/\epsilon$ is shown in Fig.3.

\section{Nuclear spin lattice relaxation rate}
In the superclean limit T$_1^{-1}$  at low temperature is given by 
\begin{eqnarray}
T_1^{-1}(T,H)/T_{1n}^{-1} &=&
\frac{1}{2T}\int_0^{\infty} 
(\frac{N(E,H)}{N_0})^2
{\rm sech}^2(\frac{E}{2T})
\nonumber\\
&=& (\frac4\pi \frac\epsilon\Delta)^2 J(T/\epsilon)
\end{eqnarray}
where
\begin{equation}
J(T/\epsilon)=
\frac{\epsilon}{2T}
\int_0^{\infty}ds g^2(s)
{\rm sech}^2(\frac{\epsilon}{2T}s)
\end{equation}             
Here $T_{1n}^{-1}$  is the nuclear spin lattice relaxation rate
in the normal state. 
We have 
$\displaystyle J(T/\epsilon) \rightarrow
1$ for $T/\epsilon
\ll 1$ and
$\displaystyle J(T/\epsilon)\rightarrow \frac13(\frac{\pi^2 T}{4\epsilon})^2 $
for $T/\epsilon \gg 1$.
Therefore the nuclear spin lattice 
relaxation rate $T_1^{-1}(T,H)/T_{1n}^{-1}$
takes 
$\sim H$
behavior in the limit of $T \rightarrow 0$ 
and  $\sim (\frac{T}{\Delta})^2$  in the limit of  
$T \gg \epsilon$
Again it is more convenient to introduce the scaling function by
\begin{eqnarray}
G(T/\epsilon) \equiv
\frac{T_1^{-1}(T,H) - T_1^{-1}(T,0)}
{T_{1n}^{-1} [\frac4\pi \frac{\epsilon}{\Delta}]^2}
= J(T/\epsilon) - \frac{\pi^4}{48} (T/\epsilon)^2 \nonumber\\
\simeq
\left\{
\begin{array}{ll}
1 -(\frac{\pi^4}{12} -\frac{\pi}{36}) (\frac{T}{\epsilon})^2 + \cdots
& \mbox{for $\frac{T}{\epsilon} \ll 1$} \\
\frac{\pi^2}{16} + O((\frac{\epsilon}{T})^2) 
&
\mbox{for $\frac{T}{\epsilon} \gg 1$}
\end{array}
\right.
\end{eqnarray}

We show the scaling function $G(T/\epsilon)$ 
in Fig.3.
The detection of this scaling function will be very useful.

\section{Planar thermal conductivity}
Following \cite{won_nato} 
the thermal conductivity in the superclean limit 
(i.e.$\Gamma/\Delta \ll
H/H_{c2} \ll 1$ )
within the conducting plane in the 
vortex state with the field configuration of ${\bf H} \parallel {\bf c}$
 is obtained as 
\begin{equation}
\kappa(T,H)/\kappa_n
=\frac{56}{5\pi}(\frac{T}{\Delta})^2
\{ \frac{20}{7\pi^2}(\frac{\epsilon}{T})^2
+ (\ln [\frac{4\Delta}{3.5 T\sqrt{1+ (\frac{\epsilon}{1.75T})^2}}
])^2 \}
\end{equation}
where $\kappa_n =\frac{\pi^2}{6}\frac{n T}{\Gamma m}$ is 
 the thermal conductivity in the
normal state, $n$ the density of  electrons 
$m$ the electron mass, and $\Gamma$ the scattering rate by impurity.
In the limit of $T \rightarrow 0$,
$\kappa(T,H)/\kappa_n$ reduces to  
$\displaystyle \frac{8v^2}{\pi^2 \Phi_0 \Delta^2} H$,
linear in $H$ contrary to \cite{kubert}.
This linear field dependence as well as the quasi-scaling behavior
of $\kappa/T^3$ versus $H/T^2$ in Eq.(14) have not 
been seen in high-$T_c$ cuprates.
\par 
On the other hand, a recent thermal conductivity data of 
Sr$_2$RuO$_4$ at $T=$ 0.35 K
exhibits a clear $H$-linear dependence for $H> 0.02$ T \cite{izawa}.
At a low magnetic field ($H < 0.015$ T) the thermal conductivity
appears to be independent of $H$.
The thermal conductivity in the superclean limit
in the present model for $T> \epsilon$ becomes almost independent of
$H$ and increases like $\kappa \sim T^3 \ln (\Delta /T)$.
\par
Further, this kind of
behavior should be quite common to 
the unconventional superconductors with the nodal superconductors
(e.g. E$_{2u}$-state in UPt$_3$ \cite{suderow}).
In particular when $\epsilon/T \ll 1$, $\kappa/T^3$ 
decreases with increasing $H$ almost linearly in 
$H/T^2$.
Indeed this behavior has been observed in the
B-phase of UPt$_3$ recently \cite{suderow}.
\par
The low temperature thermal conductivity in high-$T_c$ cuprates
with the field ${\bf H} \parallel {\bf c}$ appears
to be described by the one in the clean limit
(i.e. $H/H_{c2} \ll \Gamma/\Delta \ll 1 $ )
rather than the one in the superclean limit.
\par 
The thermal conductivity in the clean limit,
on the other hand, is given by \cite{won_nato};
\begin{equation}
\kappa/\kappa_{00}
= \frac{\Delta_{00}}{\Delta}
\big( 1 + \frac{v^2 e H}{6 \pi \Gamma \Delta}
\ln (4 \sqrt{\frac{2\Delta}{\pi\Gamma}})
\ln ( \frac{4\Delta}{v\sqrt{e H}}) \big)
\end{equation}
Here $\Delta_{00}$ is the order parameter
in the absence of the impurity scattering at
$T=0$, $\kappa_{00}=$  
$ \frac\pi3 \frac{T n}{\Delta_{00} m}$
the 
universal thermal conductivity in the limit of
$T \rightarrow 0$ and for $H=$ 0.   
Indeed this $ \sim H \ln (const./H)$ behavior 
is very consistent with the recent data from YBCO
and Bi2212 single crystals \cite{chiao}.
We show such a comparison in Fig.4, where we took 
$\Gamma/\Gamma_c = $ 0.06, 0.188, and 0.33. 
Here $\Gamma_c=$ 0.8819 $T_{c0}$ is
the critical scattering rate when the 
superconductivity disappears.

\begin{figure}
\caption{
The thermal conductivity data in YBCO by Chiao {\it et al.}
[26] are fitted with the the thermal conductivity 
Eq.(15) with
$\Gamma/\Gamma_c =$  0.06(----),
0.188($-$ $-$ $-$), 0.33($\cdots$),  where  $\Box$ for $x=$ 0,
$\Diamond$ for $x=$ 0.006, $\triangle$ for $x=$ 0.03 of 
YBa$_2$(Cu$_{1-x}$Zn$_x$)$_3$O$_{6.9}$.
The $H \ln(const./H)$ dependence describes very well the 
experimental data.
}
\end{figure}

\section{Concluding Remarks}
Limiting ourselves in the configuration ${\bf H} \parallel {\bf
c}$ (i.e. the magnetic
field normal to the conducting plane) and in the superclean limit
we obtain the expression of the thermodynamic quantities and $T_1^{-1}$
in NMR in vortex states in $d$-wave superconductors. Some of
limiting behaviors for $T \rightarrow 0$ K have been well established in the
vortex state of YBCO, though little work on the scaling behavior
has been done. 
We have shown also the
present model describes some features of the vortex state in
Sr$_2$RuO$_4$, specially the presence of the nodal structure
in the order parameter.
This suggest that single crystals of 
Sr$_2$RuO$_4$ will provide another testing ground
of the present model.
At the present,
among three 2D $f$-wave states for 
Sr$_2$RuO$_4$, only one 
of them, $\Delta({\bf k}) \sim \cos(ck_3) e^{\pm i\phi}$,
 appears to be viable,
since both the extremely small angle dependence of 
the upper critical field \cite{mao}
and the thermal conductivity \cite{izawa,tanatar}
with the field applied parallel to the conducting plane
are incompatible to other two candidates.
Also, we expect similar behaviors of vortex state
in other unconventional superconductors \cite{won_nato,won_press}. 
Therefore the
exploration of the vortex state will bring further insight in
unconventional superconductors. 

\section{Acknowledgment}
First of all we thank May Chiao for
providing us her Ph.D. thesis, 
which was very useful for the present work.
Our thank also goes to Bernard Revas and
Yuji Matsuda, who send us
Ref.\cite{wang}and Ref.\cite{izawa}, respectively, prior to 
the publication.
Finally we 
thank Thomas Dahm and  Koichi Izawa for useful
discussions on the related subjects.
HW acknowledges the support from the
Korean Science and Engineering Foundation 
through the Grant No. 1999-2-114-005-5.
Also, HW thanks Dept. of Physics and Astronomy, USC
for their hospitality during her stay.

\newpage

\begin{minipage}[t]{.40\linewidth}
\vskip 8cm
\centering
\psfig{figure=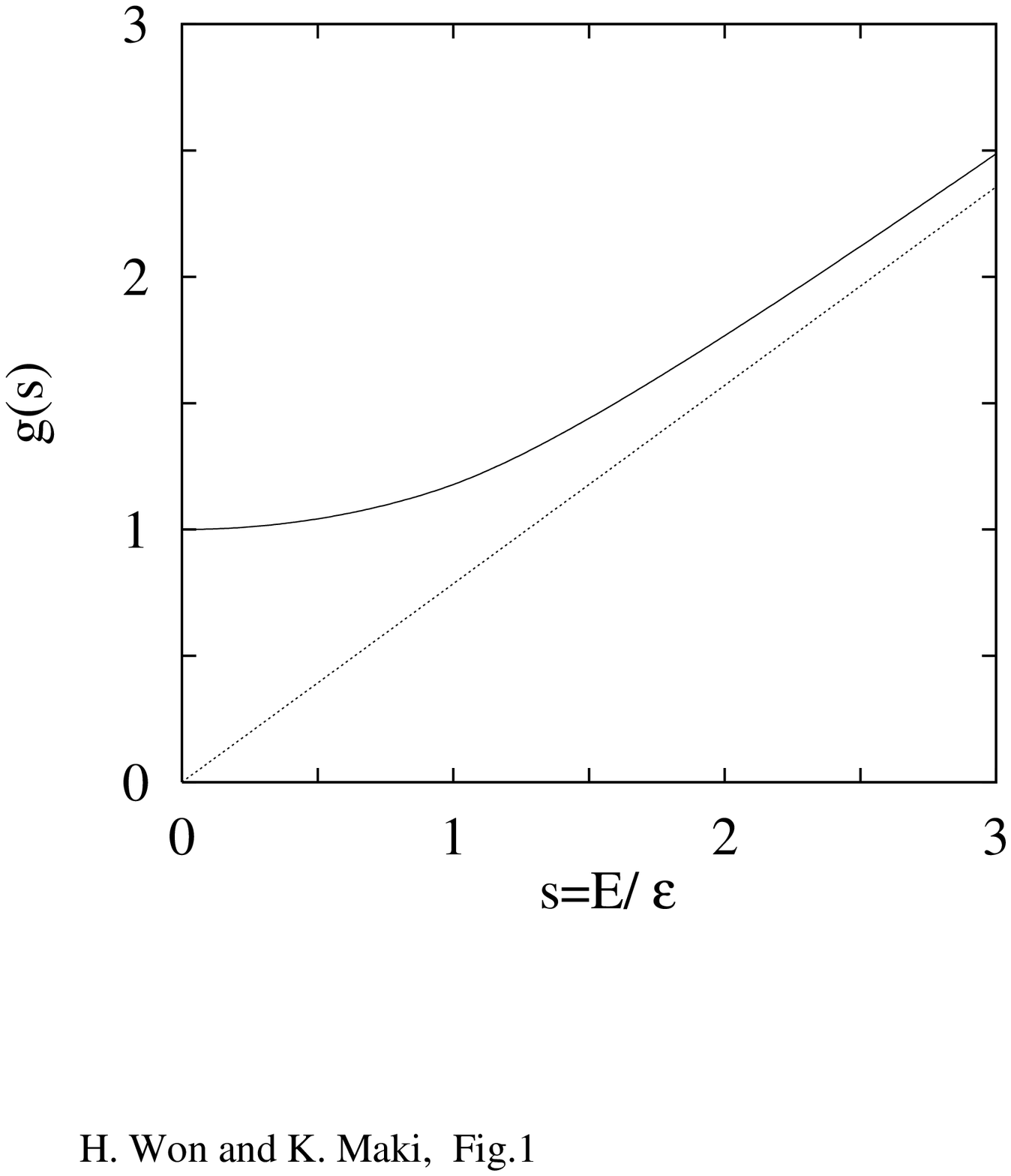}
\end{minipage}

\begin{minipage}[t]{.40\linewidth}
\vskip 8cm
\centering
\psfig{figure=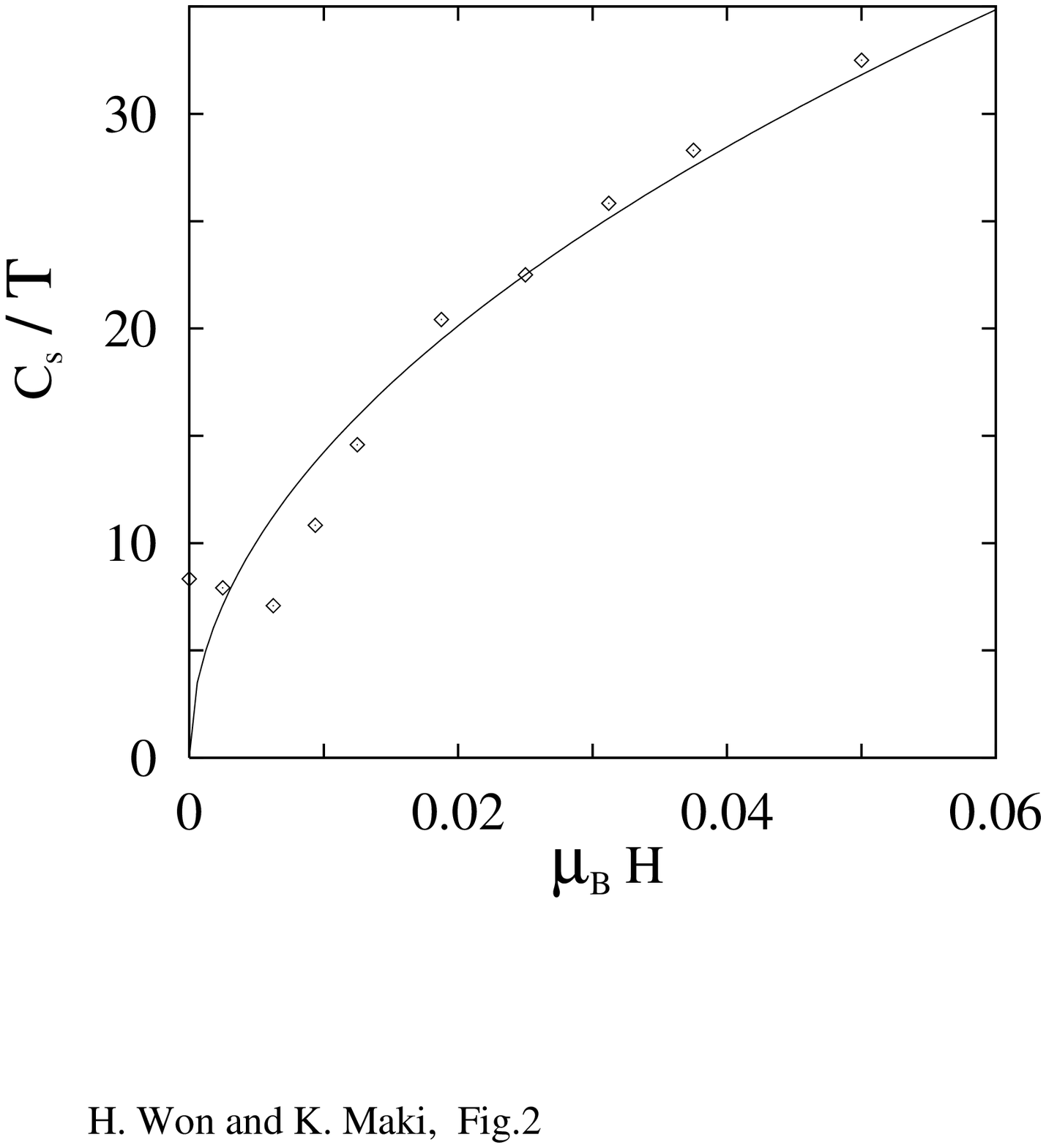}
\end{minipage}

\begin{minipage}[t]{.40\linewidth}
\vskip 8cm
\centering
\psfig{figure=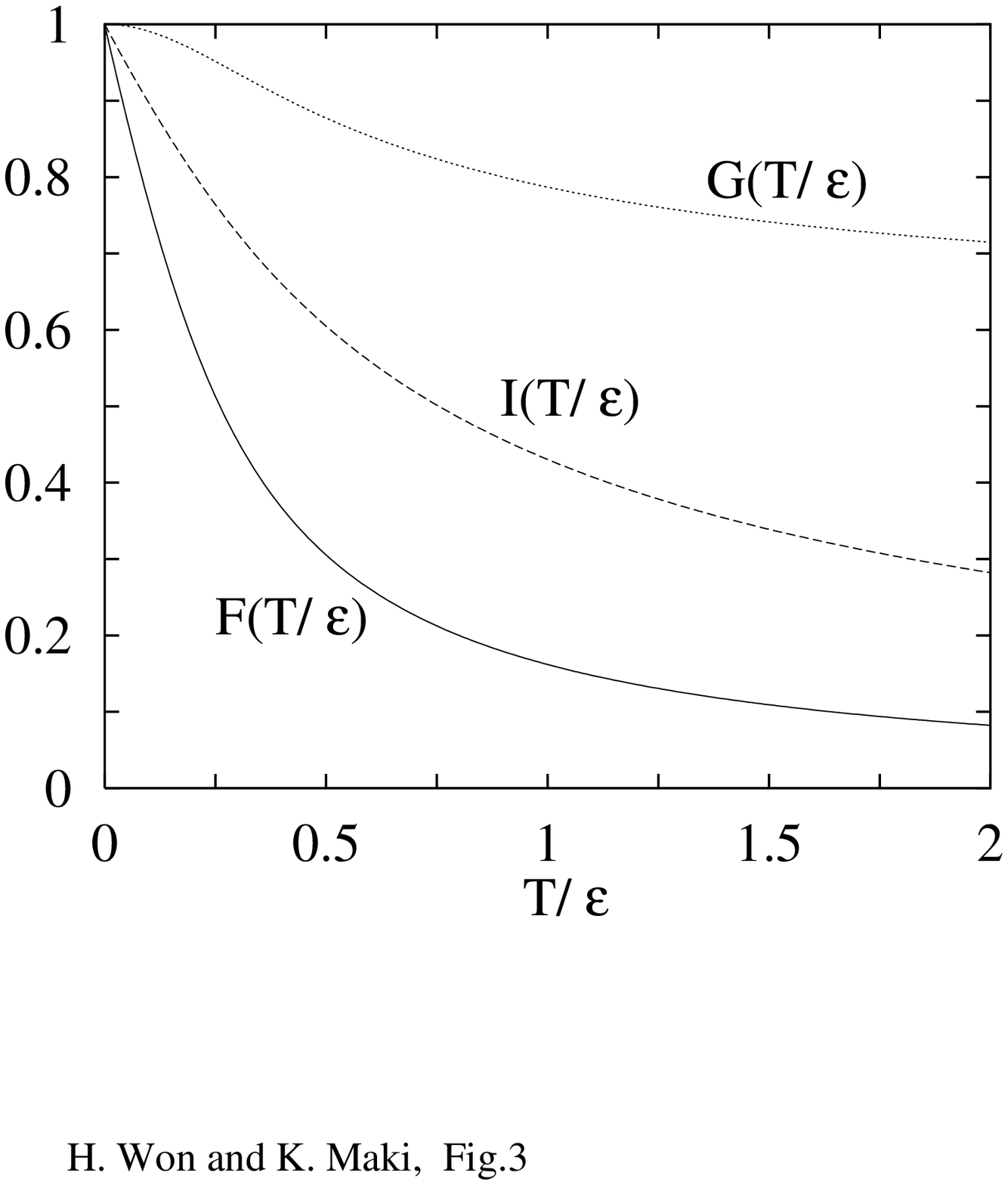}
\end{minipage}

\begin{minipage}[t]{.40\linewidth}
\vskip 8cm
\centering
\psfig{figure=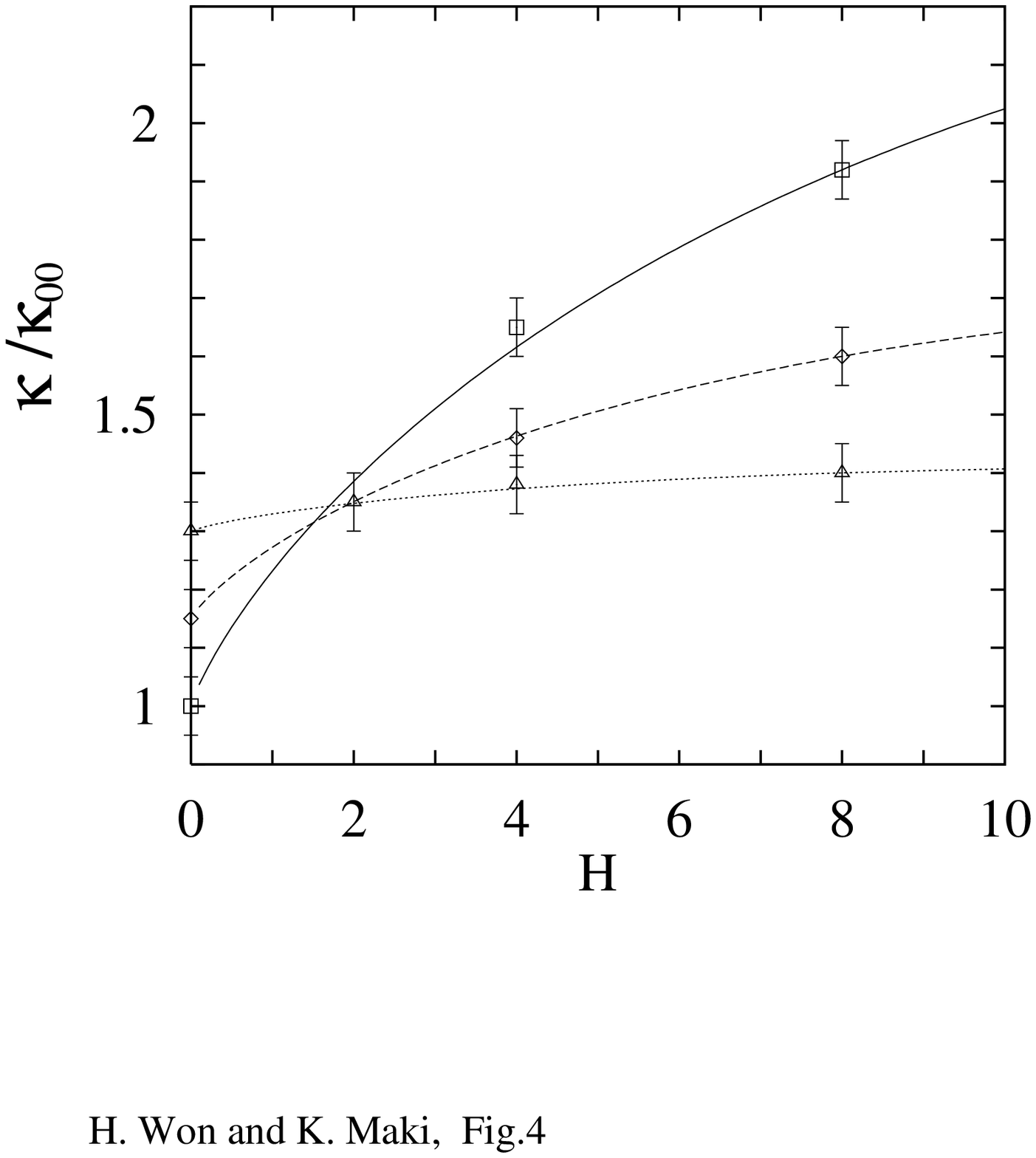}
\end{minipage}

\end{document}

%% file: euromacr.tex
\def\And{{\rm and\ }}

\newif\ifboo \boofalse
